\documentclass[aps,prl,twocolumn,showpacs,superscriptaddress,floatfix,nofootinbib]{revtex4}
\usepackage{graphicx}
\usepackage{amsmath}
\usepackage{epsfig}
\usepackage{helvet}
\usepackage{amssymb}

\newcommand{\be}{\begin{equation}}
\newcommand{\ee}{\end{equation}}
\newcommand{\bea}{\begin{eqnarray}}
\newcommand{\eea}{\end{eqnarray}}

\newcommand{\tr}{\mbox{tr}}
\newcommand{\bra}[1]{\mbox{$\langle #1 |$}}
\newcommand{\ket}[1]{\mbox{$| #1 \rangle$}}

\newcommand{\proj}[1]{\mbox{$|#1\rangle \!\langle #1 |$}}

\def\V{\mathbb{V}}

\def\S{{\mathbb S}}

\def\LLL{{\cal L}}
\def\B{{\cal B}}
\def\tr{ \mbox{tr}}
\def\R{{\cal R}}

\begin{document}

\title{
Entanglement renormalization }

\author{G. Vidal}
\affiliation{School of Physical Sciences, the University of
Queensland, QLD 4072, Australia} \affiliation{Institute for Quantum
Information, California Institute for Technology, Pasadena, CA 91125
USA}
\date{\today}

\begin{abstract}

In the context of real-space renormalization group methods, we propose a novel scheme for quantum systems defined on a D-dimensional lattice. It is based on a coarse-graining transformation that attempts to reduce the amount of entanglement of a block of lattice sites before truncating its Hilbert space.
Numerical simulations involving the ground state of a 1D system at criticality show that the resulting coarse-grained site requires a Hilbert space dimension
that does not grow with successive rescaling transformations. As a result we can address, in a quasi-exact way, tens of thousands of quantum spins with a computational effort that scales logarithmically in the system's size. The calculations unveil that ground state entanglement in extended quantum systems is
organized in layers corresponding to different length scales. At
a quantum critical point, each rellevant length scale makes an equivalent
contribution to the entanglement of a block with the rest of the system.

\end{abstract}


\maketitle

Renormalization, one of the conceptual pillars of statistical
mechanics and quantum field theory, revolves around the idea of
rescaling transformations of an extended system \cite{fisher}. These so-called renormalization group (RG) transformations are not only a key
theoretical element in the modern formulation of critical phenomena
and phase transitions, but also the basis of important computational methods for many-body problems.

In the case of quantum systems defined on a lattice, Wilson's real-space RG methods
\cite{wilson}, based on the truncation of the Hilbert space,
replaced Kadanoff's spin-blocking ideas \cite{kadanov} with a
concise mathematical formulation and an explicit prescription to implement
rescaling transformations. But it was not until the advent of White's
density matrix renormalization group (DMRG) algorithm \cite{dmrg} that RG methods became the undisputed numerical approach for systems on a 1D lattice. More recently such techniques have gained renewed momentum under the influence of
quantum information science. By paying due attention to
entanglement, algorithms to simulate time-evolution in 1D systems \cite{visim} and to address 2D systems \cite{veci} have been put forward.

The practical value of DMRG and related methods is unquestionable. And yet, they notably fail to satisfy one of the most natural requirements for a RG transformation, namely to have scale invariant systems as fixed points. Instead, for such systems the number of degrees of freedom (that is, the Hilbert space dimension) of an effective site increases with each rescaling transformation. This fact conflicts with the very spirit of renormalization, but it is not only disturbing from a conceptual viewpoint. The cost of a computation depends on the size of the effective sites and it becomes unaffordable after a sufficiently large number of iterations.

In this Letter we propose a real-space RG transformation for quantum systems on a D-dimensional lattice that, by {\em renormalizing} the amount
of entanglement in the system, aims to eliminate the growth of the
site's Hilbert space dimension along successive rescaling transformations. In particular, when applied to a scale invariant system the
transformation is expected to produce a coarse-grained system
identical to the original one. Numerical tests for critical systems in $D=1$ spatial
dimensions confirm this expectation. Our results also unveil the
stratification of entanglement in extended quantum systems, and open
the path to a very compact description of quantum
criticality.

\begin{figure}
  \includegraphics[width=8cm]{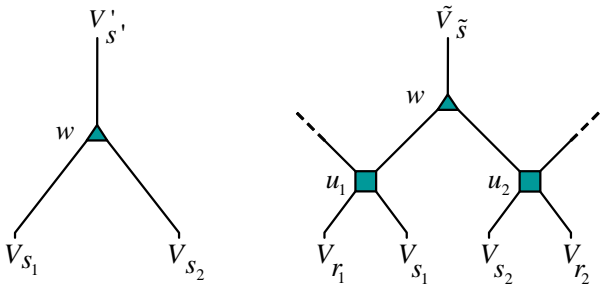}\\
  \caption{
Isometries and disentanglers. Left: a standard numerical RG
transformation builds a coarse-grained site $s'$, with Hilbert space
dimension $m$, from a block of two sites $s_1$ and $s_2$ through the
isometry $w$ of Eq. (\ref{eq:isometry}). Right: by using the
disentanglers $u_1$ and $u_2$ of Eq. (\ref{eq:disentanglers}),
short-range entanglement residing near the boundary of the block is
eliminated before the coarse-graining step. As a result, the
coarse-grained site $\tilde{s}$ requires a smaller Hilbert space dimension
$\tilde{m}$, $\tilde{m} < m$.
   }\label{fig1}
\end{figure}

{\bf Real-space RG and DMRG.--}
As originally introduced by Wilson \cite{wilson}, real-space RG
methods truncate the local Hilbert space of a block of sites in order to reduce its degrees of freedom.
Let us consider
a system on a lattice $\LLL$ in D spatial dimensions and its Hilbert space
\begin{equation}\label{eq:lattice}
    \V_{\LLL} \equiv \bigotimes_{s \in \LLL} \V_{s},
\end{equation}
where $s \in \LLL$ denotes the lattice sites and $\V_{s}$ has finite
dimension. Let us also consider a block $\B \subset \LLL$ of neighboring sites,
with corresponding Hilbert space
\begin{equation}\label{eq:block}
    \V_{\B} \equiv \bigotimes_{s \in \B} \V_{s}.
\end{equation}
In an elementary RG transformation, the lattice $\LLL$ is mapped
into a new, {\em effective} lattice $\LLL'$, where each site
$s'\in \LLL'$ is obtained from a block $\B$ of
sites in $\LLL$ by {\em coarse-graining}. More specifically, the space $\V'_{s'}$ for
site $s'\in \LLL'$ corresponds to a subspace $\S_{\B}$ of $\V_{\B}$,
\begin{equation}\label{eq:coarse-grained}
    \V'_{s'} \cong \S_{\B} \subseteq \V_{\B},
\end{equation}
as characterized by an isometric tensor $w$, Fig (\ref{fig1}),
\begin{equation}\label{eq:isometry}
    w:\V'_{s'} \mapsto \V_{\B},~~~~~ w^{\dag}w = I.
\end{equation}
Isometry $w$ can now be used to map a state $\ket{\Psi}\in \V_{\LLL}$ [typically, a ground state] into a coarse-grained state $\ket{\Psi'}\in \V_{\LLL'}$. 
A thoughtful selection of subspace $\S_{\B}$ is essential. 
On the one hand, its dimension $m$ should be as small as possible, because the cost of subsequent tasks, such as computing expectation values for local observables, grows polynomially with $m$. On the other, $\S_{\B}$ needs to be large enough that $\ket{\Psi'}$ retains all relevant properties of $\ket{\Psi}$. White identified the optimal choice as part of his DMRG algorithm \cite{dmrg}. Let $\rho^{[\B]}$ denote the reduced {\em density matrix} of $\ket{\Psi}$ on block $\B$. Then,
\begin{equation}\label{eq:truncation}
   \S_{\B} \equiv <\ket{\rho_1}, \cdots, \ket{\rho_m}>,
\end{equation}
where $\ket{\rho_i}$ are the $m$ eigenvectors of $\rho^{[\B]}$ with largest eigenvalues $p_i$, and $m$ is such that $\epsilon \geq 1-\sum_{i=1}^m p_i$, where $\epsilon$ is a preestablished truncation error, $\epsilon \ll 1$. 

Notice that $m$ refers to the amount of
entanglement between $\B$ and the rest of the lattice, $\LLL-\B$, as
characterized by the rank of the truncated Schmidt decomposition
\begin{equation}\label{eq:Schmidt}
    \ket{\Psi} \approx \sum_{i=1}^{m} \sqrt{p_i}
    \ket{\rho_i}\otimes \ket{\sigma_i},~~~~~\ket{\sigma_i} \in \V_{\LLL-\B}.
\end{equation}
In other words, the performance of DMRG-based methods depends on the amount of
entanglement in $\ket{\Psi}$.

{\bf Entanglement Renormalization.--} We propose a technique to
reduce the amount of entanglement between the block $\B$ and the
rest of the lattice $\LLL$ while still obtaining a quasi-exact
description of the state of the system. This is achieved by deforming, by means of a unitary transformation, the
boundaries of the block $\B$ before truncating its Hilbert space, see Fig.
(\ref{fig1}).

Let us specialize, for simplicity, to a 1D lattice and to a block $\B$ made of just two contiguous sites
$s_1$ and $s_2$. Let $r_1$ and $r_2$ be the two sites immediately to
the left and to the right of $\B$. Then we consider unitary transformations $u_1$ and $u_2$, the {\em disentanglers},
acting on the pairs of sites $r_1s_1$ and $s_2r_2$,
\begin{eqnarray}\label{eq:disentanglers}
    u_1:\V_{r_1}\otimes \V_{s_1} \rightarrow \V_{r_1}\otimes \V_{s_1},
     &&u_1^{\dag}u_1 = u_1u_1^{\dag} = I, \nonumber\\
    u_2:\V_{s_2}\otimes \V_{r_2} \rightarrow \V_{s_2}\otimes \V_{r_2},
     &&u_2^{\dag}u_2 = u_2u_2^{\dag} = I.
\end{eqnarray}
Properly chosen disentanglers reduce the {\em
short-range} entanglement between the block $\B$ and its immediate neighborhood. The original state $\rho^{[\B]}$ of block $\B$,
\begin{equation}\label{eq:entangled_rho}
    \rho^{[\B]} = \tr_{r_1r_2} \left[\rho^{[r_1s_1s_2r_2]}\right],
\end{equation}
is now replaced with a partially disentangled state $\tilde{\rho}^{[\B]}$,
\begin{equation}\label{eq:disentangled_rho}
    \tilde{\rho}^{[\B]} = \tr_{r_1r_2} \left[(u_1\otimes u_2)
    \rho^{[r_1s_1s_2r_2]} (u_1\otimes u_2)^{\dag}\right],
\end{equation}
which has a smaller effective rank $\tilde{m}$, $\tilde{m} <
m$. Our RG transformation consists of two steps:

$(i)$ First we decrease or {\em renormalize} the amount of entanglement between block
$\B$ and the rest of $\LLL$. This is achieved with disentanglers that act locally around the boundary of $\B$. The block does not become completely disentangled because only entanglement localized near its boundaries can be removed.

$(ii)$ Then, as in Wilson's proposal, we {\em truncate} the Hilbert
space of block $\B$, and we do so following White's idea to target
the support of the block's density matrix. But instead of keeping
the support of the original $\rho^{[\B]}$, we retain the (smaller) support of the
partially disentangled density matrix $\tilde{\rho}^{[\B]}$.

\begin{figure}[h]
  \includegraphics[width=8.5cm]{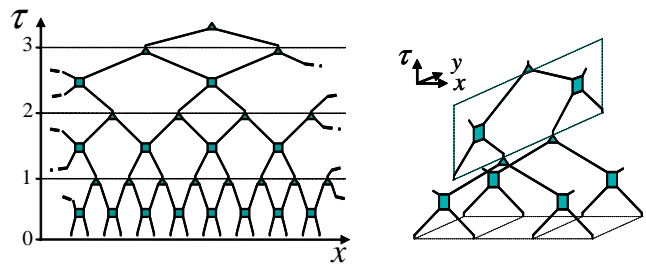}\\
  \caption{Left: MERA for a 1D lattice with periodic boundary
conditions. Notice the fractal nature of the tensor network.
Translational symmetry and scale invariance can be naturally
incorporated, substantially reducing the computational complexity of
the numerical simulations. Right: building block of a MERA for a
2D lattice. Disentanglers and isometries address one of the $x$ and $y$ spatial
directions at a time.
  }\label{fig2}
\end{figure}

The above steps, characterized by unitary and isometric tensors $u$ and $w$, produce an alternative coarse-grained lattice $\tilde{\LLL}$ and a coarse-grained state $\ket{\tilde{\Psi}}\in \V_{\tilde{\LLL}}$ that contains less entanglement than $\ket{\Psi'}\in \V_{\LLL'}$. Importantly, the expectation value 
	$\bra{\Psi}o\ket{\Psi}=\tr\left[ o\rho^{[\R]}\right]$,
where $o$ is an observable defined on a small set of sites $\R\subset\LLL$, can be efficiently computed from just $\ket{\tilde{\Psi}}$, $u$, $w$ and $o$. There are two ways:  ($i$) as described in \cite{MERA}, $\rho^{[\R]}$ can be computed from $\tilde{\rho}^{[\tilde{\R}]}\equiv \tr_{\tilde{\LLL}-\tilde{\R}}\proj{\tilde{\Psi}}$, where $\tilde{\R}$ is the set of sites of $\tilde{\LLL}$ causally connected to $\R$ through $u$ and $w$; ($ii$) alternatively, we can use $u$ and $w$ to {\em lift} $o$ from $\LLL$ to $\tilde{\LLL}$, producing $\tilde{o}$, and exploit that $\bra{\Psi}o\ket{\Psi} = \bra{\tilde{\Psi}}\tilde{o}\ket{\tilde{\Psi}}$. 

The lifting of linear operators from $\LLL$ to $\tilde{\LLL}$ generates, by iteration, a well-defined RG flow in the space of hamiltonians such that, in the 1D case, an interaction term $h$ acting on at most three consecutive sites of $\LLL$
is mapped into an interaction term $\tilde{h}$ acting also on at most three consecutive sites in $\tilde{\LLL}$ [analogous rules apply in
$D>1$ spatial dimensions]. Thus, in spite of the fact that disentanglers deform the original tensor product structure of $\V_{\LLL}$, the above rescaling transformation preserves {\em locality}, in that it maps local theories into local theories.

\begin{figure}
  \includegraphics[width=8cm,height=8cm]{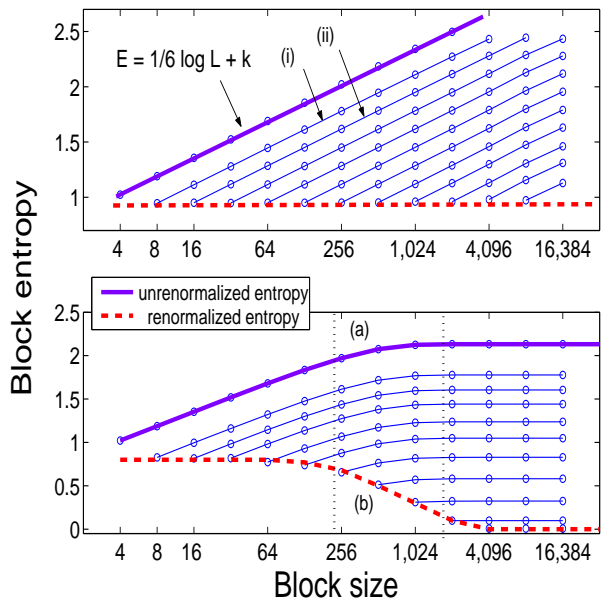}\\
  \caption{
Scaling of the entropy of entanglement in 1D quantum Ising model
with transverse magnetic field. Up: in a
critical lattice [$h=1$ in Eq. (\ref{eq:ising_ham})], the
unrenormalized entanglement of the block scales with the block size
$L$ according to Eq. (\ref{eq:critical}). Instead, renormalized
entanglement remains constant along successive RG transformations,
as a clear manifestation of scale invariance. Line (i)
corresponds to using disentanglers only in the first RG
transformation. Line (ii) corresponds to using disentanglers only in
the first and second RG transformation. Down: in a noncritical
lattice [$h=1.001$ in Eq. (\ref{eq:ising_ham})], the unrenormalized
entanglement scales roughly as in the critical case until it
saturates (a) for block sizes comparable to the correlation length. Beyond that length scale, the renormalized entanglement
vanishes (b) and the system becomes effectively unentangled.
  }\label{fig3}
\end{figure}

In the same way as DMRG is related to matrix product states (MPS)
\cite{mps}, the above RG transformation is naturally associated to a
new ansatz for quantum many-body states on a $D$ dimensional
lattice. The {\em multi-scale entanglement renormalization ansatz}
(MERA) consists of a network of
isometric tensors (namely the isometries $u$ and disentanglers $w$ corresponding to successive iterations of the RG transformation) locally connected
in $D+1$ dimensions. The extra direction $\tau$, related to
the RG flow, grows only as the logarithm of the lattice dimensions,
see Fig. (\ref{fig2}). Several properties of the MERA together with its connection to quantum circuits are described in \cite{MERA}.

\begin{figure}[h]
  \includegraphics[width=8cm,height=8cm]{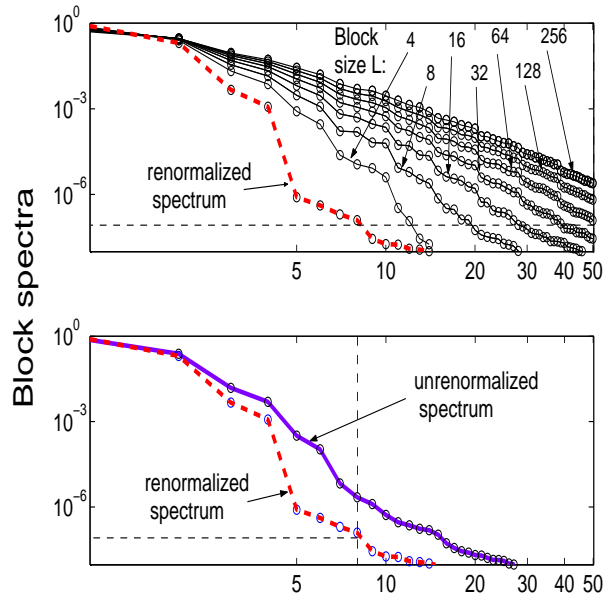}\\
  \caption{
Spectrum of the reduced density matrix of a spin block. Up: as the
size $L$ of the spin increases, the number $m$ of eigenvalues
$\{p_i\}$ required to achieve a given accuracy $\epsilon$, see Eq.
(\ref{eq:truncation}), also increases. In particular, $m$ grows
roughly exponentially in the number $\tau=\log_2 L$ of RG
transformations. The spectrum resulting from applying disentanglers
leads to a significantly smaller $\tilde{m}$ invariant along
successive RG transformations. Down: spectrum of the reduced density
matrix of $2^\tau$ spins immediately before and after using the
disentanglers at the $\tau^{th}$ coarse-graining step. These spectra
are essentially independent of the value of $\tau=1,\cdots, 14$.
  }\label{fig4}
\end{figure}

{\bf Example.--} Figs. (\ref{fig3})-(\ref{fig4}) study the ground state of the 1D
quantum Ising model with transverse magnetic field, 
\begin{equation}\label{eq:ising_ham}
    H = \sum_{<s_1,s_2>} \sigma^x_{s_1}\otimes\sigma^x_{s_2} +
    h \sum_{s} \sigma^z_{s},
\end{equation}
for an infinite lattice \cite{ERalgorithm}.
The {\em entanglement entropy} between a block $\B$ of $L$ adjacent
spins and the rest of the lattice, defined in terms of the
eigenvalues $\{p_i\}$ of $\rho^{[\B]}$ as $S(\B) \equiv -\sum_{i} p_i
\log_2 p_i$,
scales at criticality ($h=1$) with the block size $L$ as
\cite{viscal}
\begin{equation}\label{eq:critical}
S_L \approx \frac{1}{6} \log_2 L.
\end{equation}
On the other hand, off criticality $S(\B)$ grows monotonically with $L$ until
it reaches a saturation value for a block size comparable to the
correlation length in the system. Fig. (\ref{fig3}) shows the effect of disentanglers on the
entanglement entropy of a block. Most
notably, the renormalized entanglement of a block in the critical case is reduced
to a small {\em constant} value along the RG flow. Correspondingly, as Fig. (\ref{fig4}) shows, the Hilbert space dimension $\tilde{m}$ of a block of size $L=2^{\tau}$ can be
kept constant as we increase $\tau$, allowing in principle for an arbitrary number of
iterations of the scale transformation. Our calculations, involving
the reduced density matrix of up to $L=2^{14}=16,384$ spins, have
been conducted with Hilbert spaces of dimension $\tilde{m}=8$, while
keeping the truncation error $\epsilon$ at each step fixed below
$5\times 10^{-7}$. We estimate that without disentanglers an
equivalent error $\epsilon$ requires a Hilbert space of $m \approx
500-1000$ dimensions for the largest spin blocks. This exemplifies the
computational advantatges of using disentanglers.

{\bf Discussion.--} An appealing picture emerges from the above
numerical exploration. Entanglement in the ground state of the quantum spin
chain is organized in layers corresponding to different length
scales in the system. The entanglement of a given length scale can be modified by means of a disentangler that acts on a region with linear size as large as that length scale. We obserbe two situations: ($i$) Off criticality, the entanglement between a block of sites $\B$ and the rest of the lattice $\LLL$ consists roughly of contributions from layers of length scale not greater than the correlation length in the system. Therefore, after a sufficiently large number of RG transformations with disentanglers, the effective ground state becomes a product state, that is, completely disentangled. ($ii$) At criticality, instead, the entanglement between $\R$ and the rest of $\LLL$ receives {\em equivalent} contributions from {\em all length scales} (smaller than the size of $\B$) -- see \cite{MERA} for a justification of the logarithmic scaling of Eq. (\ref{eq:critical}). In this case, by applying disentanglers we obtain a coarse-grained lattice identical to the original one, including an effective Hamiltonian identical (up to a proportionality constant) to the original one \cite{Glen}. In either case, a fixed point of the RG transformation is attained after sufficiently many iterations. Non-critical theories collapse into the trivial fixed point (a product state) while critical theories correspond to a non-trivial fixed point (an entangled state expected to depends on the universality class of critical theory). We conjecture that this picture holds also for quantum lattices in $D>1$ spatial dimensions. Interestingly, a third type of fixed point is possible, in the case of a ground state with either long-range order or topological order \cite{Miguel}.

In summary, we have presented a quasi-exact real-space RG transformation that, when tested in 1D systems, produces effective sites of bounded dimension and have (scale invariant) critical systems as its non-trivial fixed points. The key feature of our approach is a local deformation of the tensor product structure of the Hilbert space of the lattice, that identifies and factorizes out those local degrees of freedom that are uncorrelated from the rest. This deformation is such that local operators are mapped into local operators, so that relevant expectation values for the original system can be efficiently computed from the coarse-grained system.

We conclude with two remarks. First, a scale invariant critical ground state leads to disentanglers $u$ and isometries $w$ that are the same at each iteration of the RG transformation. A single pair ($u,w$), depending on $O(\tilde{m}^4)$ parameters, is thus seen to specify the ground state, leading to an extremely compact characterization that deserves further study. Second, renormalization group techniques are extensively used also in classical problems, where the MERA can efficiently represent partition functions of lattice systems, extending the scope of this work beyond the study quantum systems.

The author appreciates conversations with I. Cirac,
J. I. Latorre, T. Osborne, D. Poulin, J. Preskill and, very specially, with F. Verstraete, whose advise was crucial to
find a fast disentangling algorithm. 
USA NSF grant no EIA-0086038 and an Australian Research Council Federation Fellowship are acknowledged.



\begin{thebibliography}{99}

\bibitem{fisher} M.E. Fischer, Rev. Mod. Phys. {\bf 70}, 653 (1998).

\bibitem{wilson} See S.R. White and R.M. Noack, Phys. Rev. Lett. {\bf
68}, 3487 (1992).

\bibitem{kadanov} L. P. Kadanov, Physics {\bf 2}, 263 (1966).

\bibitem{dmrg} S. R. White, Phys. Rev. Lett. {\bf 69}, 2863 (1992), Phys. Rev. B {\bf 48}, 10345 (1993).
U. Schollwoeck, Rev. Mod. Phys. 77, 259 (2005), cond-mat/0409292.

\bibitem{visim} G. Vidal, Phys. Rev. Lett. {\bf 91}, 147902 (2003),
quant-ph/0301063; ibid. Phys. Rev. Lett. {\bf 93}, 040502 (2004),
quant-ph/0310089. S. R. White and A. E. Feiguin, Phys. Rev. Lett.
93, 076401 (2004). A. J. Daley et al, J. Stat. Mech.: Theor. Exp.
(2004) P04005, cond-mat/0403313.

\bibitem{veci} F. Verstraete, J.I. Cirac, cond-mat/0407066.

\bibitem{mps} M. Fannes, B. Nachtergaele and R. F. Werner, Comm. Math. Phys. {\bf 144}, 3 (1992), pp. 443-490.
S. \"Ostlund and S. Rommer, Phys. Rev. Lett. {\bf 75}, 19 (1995),
pp. 3537.

\bibitem{peps}
F. Verstraete, D. Porras, J. I. Cirac Phys. Rev. Lett. 93, 227205
(2004); D. Porras, F. Verstraete, J. I. Cirac, cond-mat/0504717;

\bibitem{MERA} G. Vidal, quant-ph/0610099.

\bibitem{ERalgorithm} here we have obtained an MPS for the ground state of Eq. (\ref{eq:ising_ham}) with the infinite TEBD algorithm \cite{Inf1D} and then transformed it using several rows disentanglers and isometries. An algorithm to compute the ground state at each iteration of the RG transformation, as well to obtain disentanglers, will be described elsewhere. 

\bibitem{viscal} G. Vidal, J.I. Latorre, E. Rico, A. Kitaev,
Phys. Rev. Lett. 90 (2003) 227902, quant-ph/0211074. B.-Q.Jin,
V.E.Korepin, J Stat. Phys. {\bf 116}, 79-95 (2004). P. Calabrese and
J. Cardy, J. Stat. Mech. 0406 (2004) P002.

\bibitem{Glen} G. Evenbly and G. Vidal, {\em in preparation}.

\bibitem{Miguel} M. Aguado et al., {\em in preparation}.

\bibitem{Inf1D} G. Vidal, cond-mat/0605597.

\end{thebibliography}
\end{document}